\documentclass[11pt]{article}

\usepackage[T1]{fontenc}
\usepackage[utf8]{inputenc}
\usepackage{lmodern}
\usepackage{microtype}
\usepackage[letterpaper,margin=1in]{geometry}
\usepackage{amsmath,amssymb}
\usepackage{graphicx}
\usepackage{booktabs}
\usepackage{tabularx}
\usepackage{array}
\usepackage{etoolbox}
\AtBeginEnvironment{tabularx}{}
\usepackage{enumitem}
\usepackage[font=small,labelfont=bf]{caption}
\usepackage{authblk}
\usepackage{tikz}
\usetikzlibrary{positioning,arrows.meta}
\usepackage[hidelinks]{hyperref}
\usepackage{url}

\graphicspath{{figures/}}
\newcolumntype{L}{>{\raggedright\arraybackslash}X}
\setlist{itemsep=2pt,topsep=4pt}

\hypersetup{
  pdftitle={DICOMPUTE: An Economic and Privacy-Aware Framework for Distributed Artificial Intelligence Inference},
  pdfauthor={Ritesh Kakkad, Atul Khekade, Anil Chinchawale, Omkar Mestry, Behnam Mohammadkhani},
  pdfkeywords={distributed inference, heterogeneous computing, inference economics, data sovereignty, privacy, edge AI, scheduling, decentralized infrastructure}
}

\title{\textbf{DICOMPUTE: An Economic and Privacy-Aware Framework\\ for Distributed Artificial Intelligence Inference}%
\thanks{Project websites: \url{https://dicompute.ai} and \url{https://dicompute.si}.}}

\author{Ritesh Kakkad}
\author{Atul Khekade}
\author{Anil Chinchawale}
\author{Omkar Mestry}
\author{Behnam Mohammadkhani}
\affil{XDC Network Research and Engineering Team}

\date{November 2026}

\begin{document}
\maketitle

\begin{abstract}
The prevailing design of large-scale artificial intelligence inference concentrates accelerators, model weights, and request processing in centralized computing facilities. Centralization offers important benefits in accelerator utilization, high-bandwidth interconnects, operational control, and serving efficiency. Yet centralized infrastructure is not always the least-cost or most governance-compatible choice for every inference workload. This paper develops a research framework for DICOMPUTE, a heterogeneous distributed inference architecture in which policy-constrained scheduling can allocate suitable requests across privately controlled, regional, and independently operated computing resources. We derive a full-cost comparison that accounts for capital amortization, energy, network transfer, orchestration, utilization, failures, and trust-related expenses; characterize conditions under which distributed inference can outperform centralized alternatives; and specify a privacy model separating data residency, transport confidentiality, execution confidentiality, and verifiable accounting. We propose falsifiable hypotheses and a reproducible experimental design, without presenting simulated examples as measured performance. Our central conclusion is conditional: distributed inference is most promising for divisible, latency-tolerant or locality-sensitive workloads on underutilized compatible hardware, while centralized clusters retain structural advantages for tightly coupled, memory-intensive, high-throughput inference. The framework is intended to support rigorous evaluation rather than to assert universal superiority.
\end{abstract}


\section{Introduction}

Artificial intelligence infrastructure is often discussed as a choice between hyperscale data centers and small edge devices. In practice, inference demand spans a continuum: interactive language-model requests, document classification, radiology assistance, enterprise retrieval, batch embeddings, and autonomous software agents differ substantially in model size, memory footprint, privacy requirements, latency tolerance, and traffic patterns. An architecture optimized for one class need not be efficient for another.

Centralized inference can pool demand and exploit sophisticated serving engines, fast accelerator interconnects, continuous batching, and predictable operations. These advantages are substantial and must be treated as the benchmark rather than assumed away. At the same time, organizations own compatible compute that may be idle outside peak hours, while some data cannot leave an approved jurisdiction or institution. A heterogeneous inference fabric offers a way to allocate eligible workloads to available resources subject to explicit constraints.

DICOMPUTE is examined here as an architectural research case for such a fabric. This paper emphasizes generalizable systems reasoning and testable hypotheses. It is deliberately not a software implementation audit, a token promotional document, or a claim that every conceptual privacy mechanism has been deployed. The research question is not whether distribution is intrinsically superior, but when, why, and at what full cost a policy-aware distributed system is preferable to a centralized one.

The contributions are fourfold. First, we formulate an allocation problem across heterogeneous inference resources with trust and residency as hard feasibility conditions. Second, we develop an economic break-even model that distinguishes marginal cost from total cost and explicitly prices utilization and failure. Third, we define a privacy taxonomy that prevents conflation of encrypted transport with confidential execution. Fourth, we provide an empirical research protocol with matched-model baselines, uncertainty reporting, and failure-mode evaluation.

\section{Research Questions and Scope}

We consider five research questions:
\begin{itemize}
  \item \textbf{RQ1 --- Economic efficiency:} At what utilization, workload mix, and hardware ownership conditions does distributed serving reduce the fully loaded cost of a successfully completed inference?
  \item \textbf{RQ2 --- Performance:} What latency, throughput, and tail-latency penalties or benefits result from geographic and hardware heterogeneity?
  \item \textbf{RQ3 --- Sovereignty:} Can policy-aware placement ensure that sensitive payloads remain within a designated trust domain or jurisdiction?
  \item \textbf{RQ4 --- Resilience:} Does geographic and administrative diversity improve service continuity after accounting for coordination, correlated dependencies, and node churn?
  \item \textbf{RQ5 --- Accountability:} Can off-path metering and periodic settlement support auditable compensation without introducing settlement latency into inference response time?
\end{itemize}

The unit of comparison is a completed workload at a specified quality, model version, context length, service-level objective, and security requirement. Comparing only raw GPU rental rates or device purchase prices is insufficient. Training large foundation models, globally synchronous tensor parallelism over commodity internet links, and hardware-backed confidentiality without appropriate hardware are outside the demonstrated scope of this argument.

\section{Related Research and Theoretical Foundations}

\subsection{Centralized inference serving}
Modern inference servers achieve efficiency through iteration-level scheduling, continuous batching, memory management, and separation of prefill and decode resources. Orca studies iteration-level scheduling for transformer serving~\cite{orca}; vLLM introduces PagedAttention for efficient key-value cache management~\cite{vllm}. DistServe and Splitwise examine separating prompt processing from token generation to improve resource efficiency~\cite{distserve,splitwise}. These techniques may be deployed within a distributed system as well; their benefits should not be attributed exclusively to centralization.

\subsection{Distributed and memory-constrained inference}
Petals explores collaborative execution of large language models across participant devices~\cite{petals}. FlexGen investigates high-throughput generation under constrained accelerator memory through offloading~\cite{flexgen}. Edge and federated-computing research demonstrates that data locality can be valuable but also reveals coordination, heterogeneity, and privacy challenges~\cite{edge,fedavg}. These works establish that distribution is technically plausible, not that a geographically dispersed mesh will always outperform a well-utilized accelerator cluster.

\subsection{Privacy, confidential computing, and governance}
Differential privacy provides formal protection for specified released statistics or learning procedures, but does not automatically make an ordinary plaintext inference request private~\cite{dp}. Secure multiparty computation and homomorphic encryption offer stronger cryptographic possibilities at nontrivial computational and systems cost~\cite{yao,gentry}. Trusted execution environments can reduce exposure to host administrators under specific threat assumptions, but require attestation, hardware trust, and careful treatment of side channels~\cite{sgx}. For enterprise inference, locality and contractual controls may be operationally valuable even when they do not provide cryptographic confidentiality.

\section{Reference Architecture}

The proposed architecture separates a \emph{control plane}, \emph{execution plane}, and \emph{accounting plane}. The control plane authenticates requests, resolves model compatibility, evaluates placement policy, and selects eligible capacity. The execution plane hosts model runtimes on enterprise servers, regional clusters, or independent operator devices. The accounting plane records usage and periodically reconciles compensation. None of these layers inherently requires a blockchain transaction for each inference request.

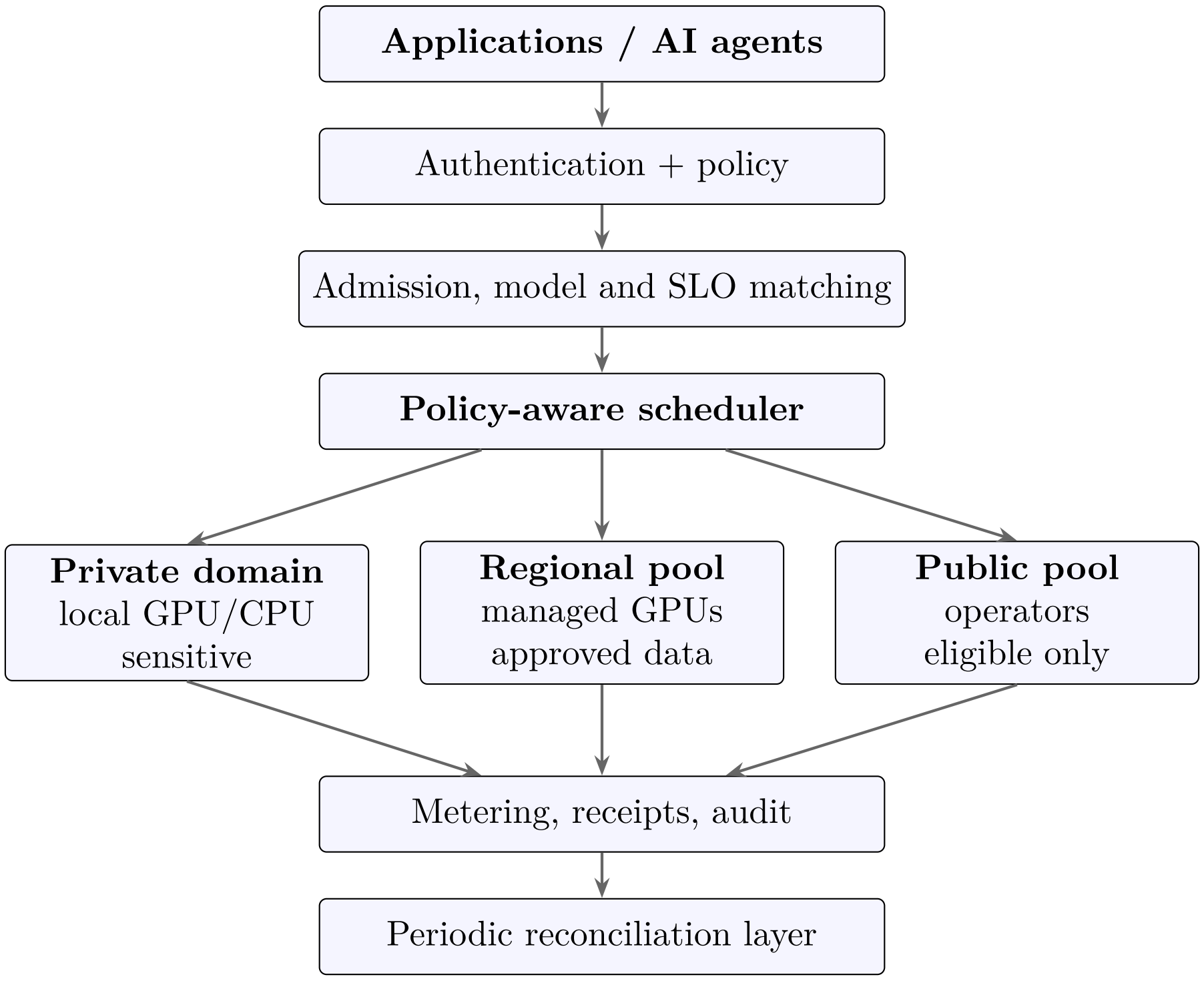
\begin{figure}[htbp]
\centering
\begin{tikzpicture}[
  box/.style={draw, rounded corners=2pt, fill=blue!4, align=center, font=\small, minimum width=5.6cm, minimum height=0.75cm},
  pool/.style={draw, rounded corners=2pt, fill=blue!4, align=center, font=\small, minimum width=3.6cm, minimum height=1.25cm},
  arr/.style={-{Stealth[length=2mm]}, thick, draw=black!60},
  node distance=0.45cm]
  \node[box] (app) {\textbf{Applications / AI agents}};
  \node[box, below=of app] (auth) {Authentication + policy};
  \node[box, below=of auth] (adm) {Admission, model and SLO matching};
  \node[box, below=of adm] (sch) {\textbf{Policy-aware scheduler}};
  \node[pool, below=0.9cm of sch] (reg) {\textbf{Regional pool}\\ managed GPUs\\ approved data};
  \node[pool, left=0.5cm of reg] (priv) {\textbf{Private domain}\\ local GPU/CPU\\ sensitive};
  \node[pool, right=0.5cm of reg] (pub) {\textbf{Public pool}\\ operators\\ eligible only};
  \node[box, below=0.9cm of reg] (met) {Metering, receipts, audit};
  \node[box, below=of met] (rec) {Periodic reconciliation layer};
  \draw[arr] (app) -- (auth);
  \draw[arr] (auth) -- (adm);
  \draw[arr] (adm) -- (sch);
  \draw[arr] (sch) -- (priv.north);
  \draw[arr] (sch) -- (reg);
  \draw[arr] (sch) -- (pub.north);
  \draw[arr] (priv.south) -- (met);
  \draw[arr] (reg) -- (met);
  \draw[arr] (pub.south) -- (met);
  \draw[arr] (met) -- (rec);
\end{tikzpicture}
\caption{Conceptual reference architecture. The figure illustrates a design space rather than asserting that every path, privacy primitive, or settlement feature is operational in a particular deployment.}
\label{fig:architecture}
\end{figure}

A privacy policy may prohibit sending any plaintext to an external worker; a model-compatibility policy may require a specific quantization, context window, or accelerator memory. In those cases the scheduler must reject incompatible nodes even when they offer a lower price. If no eligible node is available, the correct outcome is queueing, a policy-approved fallback, or rejection---not silently violating a residency rule.

\subsection{Why heterogeneous capacity matters}
The economic appeal of distributed inference is strongest when existing capacity has low incremental opportunity cost. A previously purchased enterprise GPU server that is idle overnight may be attractive for batch embedding tasks. A small local device may be suitable for compact models and low-volume private inference but unsuitable for large memory-intensive models. Model fit, tokens per second, concurrency, and memory pressure---not nominal hardware category---determine placement feasibility.

\subsection{Placement versus distributed execution}
\emph{Request-level distribution} sends independent inference requests to separate nodes. \emph{Model-parallel distribution} splits a single model execution across nodes. These are distinct engineering problems. Wide-area model parallelism incurs synchronization and communication costs and should not be assumed to work efficiently on ordinary internet connections. This paper's main economic claims concern request-level placement unless a measured experiment demonstrates otherwise.

\section{Formal Scheduling Framework}

Let $i \in I$ index inference requests and $j \in J$ index candidate nodes. Define $x_{ij} \in \{0,1\}$ to indicate placement. Let $c_{ij}$ denote expected execution cost, $n_{ij}$ network cost, $o_{ij}$ orchestration and accounting cost, $f_{ij}$ expected failure-and-retry cost, and $\ell_{ij}$ predicted end-to-end latency. Let $e_{ij} \in \{0,1\}$ indicate whether the node satisfies all privacy, jurisdiction, model, and security requirements.

A policy-constrained allocation can minimize
\begin{equation}
\min_{x} \sum_{i,j} x_{ij}\left(c_{ij} + n_{ij} + o_{ij} + f_{ij} + \lambda_i \ell_{ij}\right)
\end{equation}
subject to
\begin{equation}
\sum_{j} x_{ij} = 1, \qquad x_{ij} \le e_{ij}, \qquad \sum_{i} x_{ij}\, r_i \le K_j .
\end{equation}

Here $r_i$ is estimated resource demand, $K_j$ is available node capacity, and $\lambda_i$ converts latency into a workload-specific penalty. Real deployments additionally require time-indexed capacity, queuing dynamics, model-loading delays, and uncertainty-aware predictions. Latency can instead be expressed as a hard percentile SLO, which may be preferable to a scalar penalty in regulated workflows.

A key consequence is that \textbf{cost minimization occurs only within the feasible privacy and performance set}. Treating privacy as a soft discount risks trading away a non-negotiable requirement.

\section{Economics of Distributed Versus Centralized Inference}

\subsection{Total cost rather than headline rental rates}
For deployment $d$ over period $T$, define
\begin{equation}
\mathrm{TCO}_d(T) = A_d + E_d + N_d + O_d + S_d + F_d + R_d ,
\end{equation}
where $A$ is amortized hardware or cloud-capacity expense; $E$ energy and cooling; $N$ networking and egress; $O$ operations, orchestration and monitoring; $S$ security, compliance and audit; $F$ failures, retries and service credits; and $R$ provider compensation or capital return. These terms must be defined consistently so that provider margin is not double-counted with underlying hardware expenses.

For $Q_d$ successfully completed equivalent requests, unit cost is
\begin{equation}
U_d = \frac{\mathrm{TCO}_d(T)}{Q_d}.
\end{equation}

For text generation, the better denominator may be quality-matched output tokens or a normalized mix of input and output tokens, since prefill and decoding have different resource profiles. Report time to first token (TTFT), inter-token latency, output tokens per second, and completed requests in parallel with unit cost.

\subsection{Utilization and break-even conditions}
A simple illustrative model represents centralized cost as $C_c(Q) = F_c + v_c Q$ and distributed cost as $C_m(Q) = F_m + v_m Q$, where $F$ is fixed monthly cost and $v$ is marginal cost per normalized request. If $F_c > F_m$ but $v_c < v_m$, the break-even volume is
\begin{equation}
Q^{*} = \frac{F_c - F_m}{v_m - v_c}.
\end{equation}
Below $Q^{*}$, the distributed arrangement is cheaper; above it, the centralized one is cheaper. The conclusion reverses if relative fixed and marginal costs reverse. This model does not imply that a distributed operator has zero capital cost: ownership, depreciation, downtime, insurance, maintenance, and an economically meaningful opportunity cost must be included.

\subsection{Illustrative scenario, not a benchmark}
Assume purely for sensitivity analysis that centralized monthly fixed cost is \$6{,}000 and variable cost is \$0.0004 per normalized request, whereas a distributed pool incurs \$1{,}500 fixed and \$0.0012 variable cost. Then $Q^{*} = 5.625$ million requests per month. At one million requests, the illustrative totals are \$6{,}400 versus \$2{,}700; at ten million, \$10{,}000 versus \$13{,}500. \textbf{These values are hypothetical parameters selected to demonstrate a crossover, not DICOMPUTE measurements, vendor quotations, or forecasts.}

\begin{table}[htbp]
\centering
\small
\caption{Illustrative break-even scenario (hypothetical parameters).}
\label{tab:illustrative}
\begin{tabular}{lrrl}
\toprule
Monthly requests & Centralized (hypothetical) & Distributed (hypothetical) & Lower in illustration \\
\midrule
1 million  & \$6{,}400  & \$2{,}700  & Distributed \\
5 million  & \$8{,}000  & \$7{,}500  & Distributed \\
10 million & \$10{,}000 & \$13{,}500 & Centralized \\
\bottomrule
\end{tabular}
\end{table}

The sensitivity analysis should vary hardware depreciation, operator fees, average model size, node utilization, electricity prices, queueing overhead, failure rates, bandwidth, and data-transfer constraints. A credible study must also compare against on-demand cloud inference, which can have low fixed cost and may be preferable to either owned deployment at low volume.

\subsection{Energy and environmental impact}
Distributed deployment does not necessarily reduce energy use. Small, poorly utilized devices may consume more joules per token than consolidated accelerators; conversely, reuse of idle hardware can avoid new capital expenditure. Energy comparisons should measure wall-plug joules per completed workload and report the power-usage effectiveness assumptions of each facility. Carbon analysis requires time- and region-sensitive grid intensity and should distinguish operational emissions from embodied hardware emissions. A claim of environmental advantage without these measurements is not warranted.

\section{Privacy, Data Sovereignty, and Threat Model}

Privacy is not a single property. We distinguish:
\begin{enumerate}
  \item \textbf{Data residency:} payloads and relevant logs remain within a designated legal geography.
  \item \textbf{Administrative sovereignty:} an institution controls the operator, keys, access policy, and audit trail.
  \item \textbf{Transport confidentiality:} network intermediaries cannot read messages in transit.
  \item \textbf{Execution confidentiality:} infrastructure operators cannot access plaintext while computation occurs, under a defined adversary model.
  \item \textbf{Output and metadata protection:} generated content, prompt lengths, timing, identifiers, and accounting records are handled appropriately.
\end{enumerate}

A local inference deployment can provide meaningful residency and operational control while still exposing plaintext to administrators of the local host. Transport encryption prevents interception but does not prevent an authorized execution endpoint from seeing plaintext. Cryptographic execution confidentiality requires additional methods, such as appropriately attested confidential computing or privacy-preserving computation, with documented limitations.

\begin{table}[htbp]
\centering
\small
\caption{Execution settings, potential benefits, and residual risks.}
\label{tab:settings}
\begin{tabularx}{\textwidth}{L L L}
\toprule
Execution setting & Potential benefit & Residual risk \\
\midrule
Institutional on-premises & Locality, institutional policy control & Insider access, host compromise \\
Approved regional cluster & Geographic policy compliance & External operator access, logging \\
Independent public worker & Elastic supply, geographic diversity & Untrusted execution endpoint \\
Attested confidential worker & Stronger execution isolation under assumptions & Attestation, side channels, platform trust \\
\bottomrule
\end{tabularx}
\end{table}

\paragraph{Threat model.} Adversaries may include malicious worker operators, compromised gateways, curious infrastructure administrators, network observers, and colluding accounting participants. A defensible study must specify which adversaries each deployment resists. Model weights may also be sensitive intellectual property, and protecting them can conflict with dispatch to untrusted hosts. Encryption at rest, access controls, short-lived credentials, minimization of logs, and auditability are useful controls but are not proofs of confidential inference.

\subsection{Healthcare and regulated data}
A hospital can route identifiable imaging or clinical text only to approved infrastructure while allowing de-identified or synthetic educational tasks to use broader capacity. This creates a research opportunity for policy-preserving workload segmentation, but clinical accuracy, medical-device regulation, patient consent, institutional governance, and applicable health-data law remain independent obligations. The architecture alone does not establish regulatory compliance or clinical suitability.

\subsection{Formal policy invariant}
Let $P_i$ be the approved execution domains for request $i$, and let $D(j)$ be the domain of worker $j$. The residency invariant is
\begin{equation}
\forall i, j : \; x_{ij} = 1 \;\Rightarrow\; D(j) \in P_i .
\end{equation}
The evaluation should test whether this invariant holds during retries, worker loss, autoscaling, log collection, monitoring, and emergency fallback. A routing policy that holds only on the successful primary path is insufficient.

\section{Reliability, Tail Latency, and Failure Economics}

A distributed fleet can diversify physical failure domains, but it also introduces additional links, operator variability, software versions, and intermittent availability. Reliability must be assessed at the service level, not inferred from the number of nodes. Shared control planes, identity providers, DNS, or network dependencies can create correlated outages.

For independent retries with per-attempt failure probability $p$ and at most $k$ attempts, the simplified success probability is $1 - p^{k}$. Independence rarely holds perfectly in practice; overload, regional outages, and common software defects can correlate failures. Retry costs can also grow substantially if expensive prefill computation is repeated. The evaluation must therefore measure effective completion rate, p95/p99 latency, recovery time, and the economic cost of abandoned or duplicated work.

An effective scheduler should incorporate model warm-up time, queue length, network distance, worker reliability, and the penalty for changing execution locations. It should also expose a deterministic policy-compliant fallback path. For strict latency workloads, regional consolidation may outperform global distribution even when nominal compute prices are lower elsewhere.

\section{Deferred Metering and Economic Coordination}\label{sec:metering}

Distributed operators require credible compensation, and customers require predictable charges. Per-request on-chain transactions are generally inappropriate on the inference critical path because they introduce settlement delays, transaction costs, and dependencies unrelated to model execution. An alternative is to record authenticated usage receipts off-path, reconcile them periodically, and settle net obligations at a lower frequency.

The accounting research problem is to preserve conservation and non-duplication: an accepted request should not be billed twice; failed or cancelled requests must follow documented refund rules; and worker compensation must reconcile to validated usage. Receipts can improve auditability but do not, by themselves, prove that an untrusted worker performed a claimed computation correctly. Cryptographic proof of inference correctness is a distinct and potentially expensive research problem.

A blockchain such as XDC may serve as an optional settlement or audit anchoring layer. Its presence does not make inference private, cheaper, or verifiably correct by default. Economic claims about settlement should be evaluated separately from serving latency and model performance.

\section{Experimental Methodology}

\subsection{Comparison groups}
Evaluate at least four baselines: (B1) managed centralized cloud inference; (B2) a dedicated, optimized single-site GPU cluster; (B3) a private regional inference deployment; and (B4) a heterogeneous distributed pool with policy-aware scheduling. Use equivalent models, tokenizer versions, quantization levels, context windows, prompt sets, quality constraints, and concurrency levels wherever feasible. Report when exact equivalence is impossible.

\subsection{Workload matrix}
Use small, medium, and large models subject to actual hardware feasibility. Include interactive short prompts, long-context document analysis, retrieval-augmented generation, batched embeddings, and privacy-restricted institutional workloads. Vary request rate, burstiness, input/output token ratio, and geographic origin. A Mac mini should be tested only with models that fit its available memory and runtime; it is not a like-for-like substitute for a multi-GPU server on every model.

\begin{table}[htbp]
\centering
\small
\caption{Proposed experimental conditions.}
\label{tab:conditions}
\begin{tabularx}{0.85\textwidth}{l L}
\toprule
Dimension & Proposed experimental conditions \\
\midrule
Demand & Low, moderate, saturated, bursty \\
Model & Small / medium / large, feasible on each platform \\
Placement & Single site, approved regional, heterogeneous mesh \\
Trust & Local-only, approved operator, public eligible tasks \\
Failure & Node churn, network loss, regional disruption \\
Outputs & Cost, TTFT, p95/p99 latency, throughput, energy, compliance \\
\bottomrule
\end{tabularx}
\end{table}

\subsection{Hypotheses and statistical analysis}
\begin{description}[font=\normalfont\bfseries]
  \item[H1:] At low to moderate demand and with pre-existing eligible hardware, distributed placement lowers fully loaded cost per completed equivalent workload relative to a dedicated single-site deployment.
  \item[H2:] Under strict geographic data constraints, policy-aware scheduling achieves zero unauthorized placements in tested scenarios, including retries and failover.
  \item[H3:] At high steady-state utilization, optimized centralized inference has superior throughput or lower unit cost for at least some large-model workloads.
  \item[H4:] Deferred accounting does not materially increase user-visible inference latency compared with equivalent serving without settlement processing.
\end{description}

These are hypotheses, not findings. Run repeated trials with randomized workload order and disclose warm-up, cache state, hardware specifications, power measurements, price assumptions, and failure injection. Report confidence intervals or bootstrap uncertainty estimates; show distributions rather than only averages. Pre-register the primary metrics and decision thresholds where possible to reduce selective reporting.

\subsection{Ablation studies}
Compare cost-only routing, latency-only routing, privacy-constrained routing, and combined policy-aware routing. Disable retry optimization, node reputation, and geographic constraints one at a time to quantify their effects. Distinguish scheduling gains from improvements attributable to underlying model-serving engines such as continuous batching.

\section{Interpreting Results and Avoiding Overclaiming}

The most important expected result is a \emph{phase diagram}, not a single headline savings percentage. Its axes should include utilization, model size, latency tolerance, trust requirements, and available idle hardware. Regions of the diagram can indicate when centralization, private regional deployment, or heterogeneous placement is economically and operationally preferable.

Three common methodological errors should be avoided. First, a lower hourly hardware cost does not imply lower cost per output token if throughput is poor. Second, data residency does not imply that a worker operator cannot read plaintext. Third, synthetic engine response times do not substitute for real model execution on real hardware. A conceptual systems paper may describe a rigorous proposed framework without benchmarks, but any empirical superiority claim requires independently repeatable evidence.

\section{Applications and Broader Implications}

\paragraph{Healthcare.} Hospitals and universities may use local inference for sensitive records while allocating synthetic, anonymized, or non-sensitive educational tasks to regional resources. Research questions include policy-compliant routing and measurable local-versus-cloud cost for specific diagnostic-support workloads; clinical safety remains separate.

\paragraph{Financial services.} Banks and insurers may require strict control over customer records, transaction histories, and model logs. Policy-aware placement can support jurisdictional and institutional restrictions while reserving elastic capacity for appropriately classified tasks.

\paragraph{Public-sector sovereign AI.} Governments may benefit from geographically constrained capacity pools that include universities, agencies, and approved operators. Sovereignty depends on governance, software supply chains, model provenance, keys, and enforceable controls---not simply the physical location of servers.

\paragraph{Education and research.} Shared regional capacity can provide practical AI experimentation without requiring every institution to acquire a large accelerator cluster. Cost advantages will depend on concurrency, access rules, maintenance, and the educational workloads actually supported.

\section{Limitations and Open Problems}

The present paper offers an expanded theoretical framework and experimental protocol, not a completed multi-site performance evaluation. Its economic illustration uses hypothetical coefficients. It does not demonstrate that any particular DICOMPUTE deployment outperforms a hyperscale data center, nor that all requests are confidential from infrastructure operators. Further work is required on trustworthy worker admission, model provenance, attestation, prompt and weight leakage, heterogeneous quality assurance, adversarial receipts, settlement disputes, capacity forecasting, and energy measurement.

A broader societal question is whether distributed compute meaningfully broadens access or simply shifts cost and risk to smaller operators. Fair compensation, security responsibilities, environmental externalities, and operator concentration deserve explicit study alongside raw price comparisons.

\section{A Taxonomy of Inference Workloads and Feasible Distribution}

The viability of a distributed inference architecture depends first on what is being distributed. A single model invocation can be treated as an indivisible request, or its execution can be partitioned across devices. The two strategies have different communication costs, fault surfaces, and requirements for model state. We focus on request-level distribution because independent requests can be routed without transferring intermediate activations at each transformer layer. The appropriate unit of scheduling may be a document, an embedding batch, a conversation turn, or a complete multi-step agent task; it need not be a token.

A workload can be described by a tuple $w = (m, p, d, s, q, r)$, where $m$ is the model and its precision, $p$ the prompt and generation length distribution, $d$ the data classification, $s$ the service-level objective, $q$ the required output quality, and $r$ the residency and trust rules. A node is feasible only if its memory, runtime, model version, throughput, and governance characteristics meet all hard constraints. This feasibility test must precede price optimization. A cheap node that cannot run the model, meet the deadline, or legally process the data is not an economic alternative.

\begin{table}[htbp]
\centering
\small
\caption{Workload classes, binding constraints, and plausible placements.}
\label{tab:workloads}
\begin{tabularx}{\textwidth}{L L L L}
\toprule
Workload class & Typical binding constraint & Plausible placement & Centralized advantage \\
\midrule
Interactive long-context generation & Memory, time to first token, tail latency & Regional GPU pool or approved private cluster & Batching, large memory, fast interconnect \\
Batch embeddings and classification & Throughput, cost, deadline & Underutilized local or regional capacity & High utilization at sustained scale \\
Sensitive document assistance & Data governance, auditability & Institution-controlled inference & Managed security and mature operations \\
Agent tool selection and short responses & Bursty arrival, small model fit & Heterogeneous approved nodes & Low coordination overhead \\
Large-model tightly coupled execution & Device-to-device bandwidth & Closely coupled cluster & High-bandwidth collective communication \\
Educational sandbox inference & Budget, fairness, model availability & Shared university or regional pool & Predictable managed service \\
\bottomrule
\end{tabularx}
\end{table}

\paragraph{Implication.} A comparison that combines all these workloads into a single average price obscures the causal mechanisms of cost. Each class requires its own performance envelope and governance assumptions. In particular, medical image analysis is not equivalent to conversational inference: clinical evaluation, validation, and regulatory obligations must be addressed separately from infrastructure cost.

\subsection{Locality as an optimization resource}
Locality can reduce data transfer, improve governance, and avoid repeated movement of large documents or embeddings. But locality can also reduce pooling efficiency if each institution maintains idle capacity. The decision therefore involves a tradeoff between data movement and fragmented capacity. The relevant counterfactual is not always a remote public cloud; it may be a well-managed private regional cluster shared across institutions.

For workload $i$, let $D_i$ denote bytes transferred under a remote execution plan and $b_i$ the effective transfer price per byte, including egress, private connectivity, and security inspection. Let $K_i$ denote the local compute premium relative to a centralized alternative. Locality is financially beneficial from avoided transfer alone when $D_i b_i > K_i$, but the governance value of locality is not generally reducible to a market price. The inequality is a screening criterion, not a complete optimization rule.

\section{A Generalized Total-Cost-of-Inference Model}

A defensible cost comparison should use quality-adjusted, SLO-compliant completed inference as the denominator. Let $N$ be submitted requests, $a$ the fraction admitted, $p_s$ the probability an admitted request completes correctly and within its service objective, and $T$ the accounting horizon. Then the effective unit cost is
\begin{equation}
C_{\mathrm{effective}} = \frac{C_{\mathrm{capital}} + C_{\mathrm{energy}} + C_{\mathrm{network}} + C_{\mathrm{operations}} + C_{\mathrm{security}} + C_{\mathrm{coordination}} + C_{\mathrm{failure}}}{N\, a\, p_s}.
\end{equation}

A low advertised price can be misleading if the node experiences frequent failures, low throughput, or expensive retries. Similarly, capital equipment owned by a university is not free: a social-cost accounting perspective must include depreciation, maintenance, space, energy, and the opportunity cost of using that equipment for another task. An incremental-cost perspective may appropriately treat some of those expenses as sunk. Both viewpoints should be reported and never mixed within a single comparison.

\subsection{Capital ownership and amortization}
For node $j$, let $P_j$ be acquisition and installation cost, $R_j$ residual value, $L_j$ economic lifetime in months, and $F_j$ monthly fixed operating expense. A straight-line monthly ownership estimate is $A_j = (P_j - R_j)/L_j + F_j$. If monthly productive service capacity is $Q_j$ tokens and utilization is $u_j$, fixed cost per productive token is approximately $A_j/(u_j Q_j)$, subject to the distinction between input and output token compute. As $u_j$ approaches zero, fixed cost dominates. As $u_j$ rises, queueing delays, thermal limits, and service-level violations may increase. Thus minimum cost and maximum utilization are not equivalent objectives.

\subsection{Provider pricing versus system-wide cost}
A distributed marketplace may pay a provider less than the total resource cost or may pay more during scarcity. Provider price is a transfer between participants, whereas hardware depreciation, electricity, and operator labor are real resource expenditures. An investor-level financial analysis may optimize marketplace margin; a government-level welfare analysis may optimize total social cost and sovereignty constraints. The paper therefore recommends three parallel ledgers: payer expenditure, provider net contribution, and resource-adjusted system cost.

\subsection{Reliability-adjusted cost}
If each attempt succeeds independently with probability $p$, expected attempts to obtain one success are $1/p$. This approximation fails when outages are correlated, failures have state-dependent probabilities, or retries violate a deadline. More generally, measure $\mathbb{E}[A \mid \text{completed within SLO}]$ empirically and price failed work, duplicate execution, and recovery traffic. For requests involving sensitive data, each retry also creates another potential disclosure event unless constrained by the same policy.

\section{Threshold Analysis and Economic Phase Boundaries}

Consider a stylized centralized option with monthly fixed cost $F_c$ and variable cost $v_c$ per completed request, and a distributed option with $F_d$ and $v_d$. At volume $Q$, costs are $C_c(Q) = F_c + v_c Q$ and $C_d(Q) = F_d + v_d Q$. If $v_d > v_c$ but $F_d < F_c$, the distributed approach is cheaper only below
\begin{equation}
Q^{*} = \frac{F_c - F_d}{v_d - v_c},
\end{equation}
provided the denominator is positive and both options satisfy identical quality, security, and latency constraints. This result is algebraic, not a measured property of DICOMPUTE. It illustrates why a lower fixed-cost footprint can coexist with a higher marginal serving cost.

For illustration only, assume monthly fixed costs of 8{,}000 and 2{,}000 monetary units, with per-request variable costs of 0.0004 and 0.0010 respectively. Then $Q^{*} = 10$ million successful requests per month. Altering the assumed service capacity, prompt mix, or reliability can shift the threshold substantially. The example is deliberately dimensioned in requests rather than tokens to avoid implying that all requests have equal computational work; an empirical study should normalize by input/output token mix and model quality.

\subsection{Sensitivity to utilization}
A central facility operating near sustained optimal throughput may have lower cost per token than a heterogeneous mesh. Conversely, the mesh may offer an attractive incremental price if it activates equipment that would otherwise remain idle. A proper sensitivity study should sweep utilization jointly with workload burstiness, mean output length, queueing delay, failure rate, electricity tariff, network transfer, and operator compensation. Varying one coefficient while holding all others fixed can miss important interactions.

\subsection{The value of policy-constrained optionality}
Distributed capacity can have option value even when it is not the cheapest steady-state resource. A sovereign institution may use local capacity during network outages, compliance incidents, or temporary cloud scarcity. That option has a cost: model replication, warm capacity, testing, and standby operations. The economic question becomes whether expected avoided loss and continuity benefits justify the premium. This is an insurance-like argument and should not be conflated with routine inference savings.

\section{Scheduling Under Uncertainty}

A scheduler makes decisions using imperfect estimates of service time, queue length, network conditions, availability, and cost. An expected-cost minimum is insufficient when a small probability of a large tail delay violates the application objective. A robust formulation can minimize expected cost while imposing a chance constraint:
\begin{equation}
\Pr\left(L_i \le S_i \mid j\right) \ge 1 - \epsilon_i ,
\end{equation}
where $L_i$ is end-to-end latency, $S_i$ the request deadline, and $\epsilon_i$ an acceptable violation probability. The probability should be estimated from trace data and validated under distribution shift; assuming Gaussian latency distributions without evidence is particularly hazardous in multi-operator systems.

A practical placement hierarchy is: (1) hard legal and privacy eligibility, (2) model and hardware compatibility, (3) admission and service feasibility, (4) reliability and risk controls, and finally (5) economic optimization. This ordering prevents a weighted price function from trading away a non-negotiable legal requirement.

\subsection{Fairness and capacity concentration}
Unconstrained cost optimization may repeatedly favor a small number of providers, re-creating centralized dependence within a nominally distributed network. Provider concentration can be measured by market share, jurisdictional diversity, and correlated infrastructure dependencies. A policy might cap the fraction of sensitive traffic assigned to a single operator, but this constraint could increase cost and latency. Reporting that tradeoff is more informative than treating decentralization as a binary attribute.

\subsection{Admission control and graceful degradation}
If no eligible node can satisfy a request, a correct system should return a bounded failure, queue the task, or request a policy-approved degradation such as a smaller validated model. It should not silently reroute sensitive inputs to a public node. The evaluation should report rejection rates and timeouts as first-class outcomes; excluding them from cost calculations artificially improves apparent efficiency.

\section{A Layered Privacy and Sovereignty Framework}

Privacy claims in distributed inference often conflate four different properties. \emph{Residency} concerns where data is processed and stored. \emph{Transport confidentiality} concerns exposure while data moves. \emph{Execution confidentiality} concerns whether operators, privileged software, and other tenants can inspect plaintext during computation. \emph{Output governance} concerns logging, retention, reuse, and onward transfer. These properties can be independently present or absent.

\begin{table}[htbp]
\centering
\small
\caption{Privacy properties, example controls, and what they do not establish.}
\label{tab:privacy}
\begin{tabularx}{\textwidth}{L L L}
\toprule
Property & Example control & What it does not establish \\
\midrule
Geographic residency & Region-restricted eligible nodes & Protection from administrators in that region \\
In-transit confidentiality & Authenticated encryption & Confidentiality after decryption at the endpoint \\
Access isolation & Tenant boundaries and least privilege & Resistance to all host-level or side-channel attacks \\
Attested confidential execution & Hardware root of trust and attestation & Universal immunity to implementation defects \\
Data minimization & Local retrieval and redaction & Zero leakage from model outputs \\
Verifiable metering & Signed receipts and audit trails & Truth of every compute claim without trusted measurement \\
\bottomrule
\end{tabularx}
\end{table}

\subsection{Threat actors and attack surfaces}
The relevant actors include a malicious worker operator, compromised coordinator, dishonest accounting party, external network adversary, unauthorized tenant, insider, and compromised model supply chain. A privacy claim must name which actors it protects against. For example, keeping hospital records on an institution-owned GPU can meaningfully narrow organizational exposure, yet the institution's own privileged administrators may still have access to plaintext. A public worker with encrypted transport does not necessarily provide confidential inference.

\subsection{Sovereignty as a governance property}
A sovereign deployment requires more than physically local hardware. Dependencies on foreign model repositories, remote authentication services, control-plane availability, software updates, encryption-key custody, and payment rails can all affect operational autonomy. We propose evaluating sovereignty along five dimensions: data custody, execution control, model and software provenance, operational continuity, and institutional accountability. A jurisdiction-specific score should not replace a legal assessment; it can instead expose the assumptions that need review.

\subsection{Data lifecycle and deletion}
An end-to-end assessment should track prompts, retrieval documents, intermediate states, outputs, telemetry, crash dumps, caches, backups, and audit records. Deletion commitments should identify retention periods and enforcement mechanisms. Because some observability data may reveal personal information, logging design is part of the privacy architecture rather than an unrelated operational concern.

\section{Resilience, Energy, and Environmental Externalities}

A geographically dispersed fleet may withstand isolated hardware failures, but it can also inherit shared dependencies such as common cloud identity providers, central coordinators, software releases, or payment systems. Availability should be evaluated under both independent-node and correlated-failure scenarios. If $p_j$ is the availability of independent replicas, the probability that at least one of $k$ replicas is available is $1 - \prod_{j=1}^{k}(1 - p_j)$. This expression is an optimistic bound when failures share causes. A realistic study must include correlated regional power, network, and control-plane outages.

Energy efficiency requires similarly careful accounting. Lower capital expense does not imply lower energy use or carbon emissions. A small device may have low idle draw but poor tokens per joule for a particular model. A centralized facility may use efficient cooling and higher batching efficiency, while geographically dispersed devices may avoid additional construction and exploit existing capacity. Report energy per successful quality-equivalent output, including idle power, cooling where measurable, networking, retries, and standby capacity. Carbon comparisons additionally depend on location- and time-specific electricity emissions factors.

\subsection{Embodied infrastructure costs}
Reusing existing devices may reduce the need for new equipment, but the causal claim must consider whether the devices would otherwise be retired, repurposed, or operated at idle. Attribution of embodied emissions is sensitive to system boundaries. Report both marginal operating energy and lifecycle allocation rather than presenting one as the complete environmental footprint.

\section{Institutional Economics and Sovereign Capacity Markets}

Universities, hospitals, banks, and public agencies may possess compute resources whose utilization varies by time of day and workload. A governed capacity-sharing arrangement can potentially improve access without pooling sensitive data indiscriminately. Its viability depends on clear ownership, maintenance responsibility, model licensing, incident response, and an enforceable allocation policy. A university's spare capacity cannot be assumed to be available to a hospital merely because both are in the same country.

Three governance patterns deserve separate evaluation. In an \emph{institution-private} model, each participant retains exclusive execution control and uses a shared coordinator only for permitted metadata. In a \emph{regional trusted consortium}, participants admit approved workloads under contractual and technical rules. In an \emph{open marketplace}, operators compete to serve eligible non-sensitive work, with stronger need for reputation, verification, and fraud resistance. These patterns have different economic and security equilibria and should not be collapsed into a single deployment claim.

For a public-sector purchaser, the objective may be to minimize discounted expenditure subject to sovereignty, reliability, and workforce-development requirements. For a provider, the objective may be to maximize revenue net of electricity, maintenance, depreciation, and service risk. For a consortium, the objective may include equitable access and capacity assurance. Mechanism design must account for these competing incentives; simple per-token payments may encourage low-quality service or opportunistic underreporting unless measurement and dispute processes are credible.

\section{Experimental Design for a Publishable Evaluation}

The central empirical claim to test is not that a distributed system always wins, but that policy-aware heterogeneous placement improves cost or availability within identifiable regions of the workload space. We propose a preregistered factorial experiment with independent variables: model family and size, quantization, prompt/output length, arrival burstiness, hardware class, provider utilization, locality restriction, and failure regime.

\subsection{Baselines and matched conditions}
Compare (A) a managed centralized inference service, (B) a dedicated centralized cluster using the same model-serving engine, (C) a private institutional deployment, and (D) a heterogeneous distributed fleet. Where a baseline uses a different engine, report that as a separate comparison rather than attributing all differences to placement. Match model weights, tokenization, quality evaluation, request traces, and service-level objectives wherever technically possible.

\subsection{Primary outcomes}
Report fully loaded cost per one million input and output tokens separately, cost per SLO-compliant request, p50/p95/p99 end-to-end latency, time to first token, decode rate, successful request rate, energy per completed workload, and policy violation count. Secondary outcomes include provider concentration, retry rate, node admission latency, and settlement reconciliation delay. All metrics should include the measurement window, uncertainty intervals, and raw data provenance.

\subsection{Causal interpretation and ablations}
To isolate the contribution of scheduling, compare identical hardware fleets under cost-only, latency-only, locality-only, and joint policy-aware routing. To isolate the effect of spare capacity, compare marginal-cost and full-cost accounting on the same traces. To isolate resilience, inject node churn and common-mode control-plane failures separately. Do not infer cryptographic confidentiality from a successful residency test; security properties require their own threat-model-specific evidence.

\subsection{Statistical plan}
Run repeated trials across workload seeds and time windows. Use paired comparisons when baselines process the same request traces; report bootstrap confidence intervals for skewed cost and latency distributions. Predefine a smallest practically meaningful difference and evaluate whether confidence intervals exclude it. Publish negative findings, including workloads where centralization is superior. A complete artifact should include environment manifests, model checksums, synthetic trace generators, analysis notebooks, and raw or privacy-safe aggregated results.

\section{Discussion: A Conditional Theory of Distributed Inference}

The theoretical argument can be summarized as a constrained substitution problem. Heterogeneous compute is a useful substitute for centralized capacity only when (i) the task is executable on the available hardware, (ii) placement satisfies hard trust and jurisdictional constraints, (iii) the performance objective survives network and queueing overhead, and (iv) the full reliability-adjusted cost is competitive. The first three are feasibility conditions; the fourth is an optimization criterion. Distribution can also provide insurance against certain outages, but resilience depends on independent failure domains rather than the number of machines alone.

This theory yields several falsifiable predictions. Short, independent, locality-sensitive tasks may benefit from private or regional placement. Long-context and high-throughput generation may favor centralized accelerators with efficient batching and memory management. Extremely low-latency tasks may favor execution near the application. Highly sensitive tasks may be restricted to institution-controlled nodes even when those nodes are more expensive. The scientific contribution of a DICOMPUTE evaluation will be to measure the boundaries between these regimes, not to conceal them behind a single average savings figure.

\section{Economic Evaluation as a Research Instrument}

\subsection{What constitutes a fair comparison?}
A centralized cloud service, an owned GPU cluster, and a heterogeneous mesh are not interchangeable commodities. They differ in model availability, prompt length limits, reliability guarantees, capacity reservations, geographical constraints, and data-handling obligations. An economically meaningful comparison must therefore establish an equivalence class of admissible outputs: the same model or a validated quality-equivalent model, the same request distribution, a stated security and residency policy, and a common service-level objective (SLO). An output that is cheap but violates the SLO or privacy policy is not an equivalent substitute.

Let $x_{ij}$ denote whether request $i$ is assigned to resource $j$, and $e_{ij} \in \{0,1\}$ indicate whether the assignment meets model, residency, security, and hardware eligibility. Feasible allocations satisfy $x_{ij} \le e_{ij}$ and $\sum_j x_{ij} \le 1$. The latter inequality explicitly permits rejection; forcing every request onto some node can conceal policy violations. For accepted requests, the objective is to minimize expected successful-workload expenditure, not merely the nominal hourly price of compute.

A suitable empirical unit is one million input tokens plus a separately reported one million generated tokens, stratified by prompt and output lengths. Input and output tokens consume different resources during prefill and decode; a single undifferentiated per-token price can reverse a cost ranking when the workload mix changes. For multimodal tasks, use modality-specific units and quality criteria rather than silently equating images, audio, and text tokens.

\subsection{Fixed and variable cost frontiers}
A two-technology stylized model clarifies when savings are possible. Write $C_c(Q) = F_c + v_c Q$ for a centralized alternative and $C_d(Q) = F_d + v_d Q$ for distributed inference, where $Q$ is the number of quality-equivalent SLO-compliant requests. If $F_d < F_c$ but $v_d > v_c$, the distributed option wins below $Q^{*} = (F_c - F_d)/(v_d - v_c)$. If both the distributed fixed and marginal costs are higher, no positive-volume crossover exists; if both are lower, the distributed option dominates under the specified assumptions. These are conditional algebraic statements, not empirical findings about any deployment.

\begin{figure}[htbp]
\centering
\includegraphics[width=0.78\textwidth]{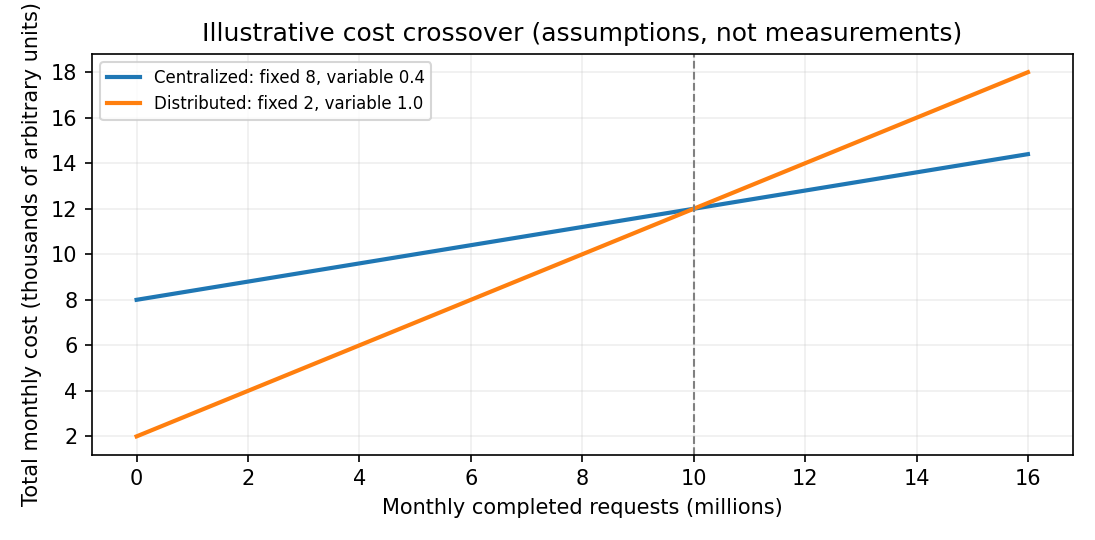}
\caption{Illustrative cost crossover: all numerical inputs are hypothetical, not measured DICOMPUTE results.}
\label{fig:crossover}
\end{figure}

The illustrative example (Figure~\ref{fig:crossover}) assumes fixed monthly costs of 8{,}000 versus 2{,}000 cost units and variable costs of 0.4 versus 1.0 thousand units per million completed requests. The resulting threshold is 10 million completed requests per month. The example intentionally does not claim that a particular GPU, Mac mini, or cloud instance has these costs. Real analysis must include taxes, financing, energy, cooling, utilization, support, replacement, egress, and quality failures.

\subsection{Utilization and the denominator problem}
\begin{figure}[htbp]
\centering
\includegraphics[width=0.78\textwidth]{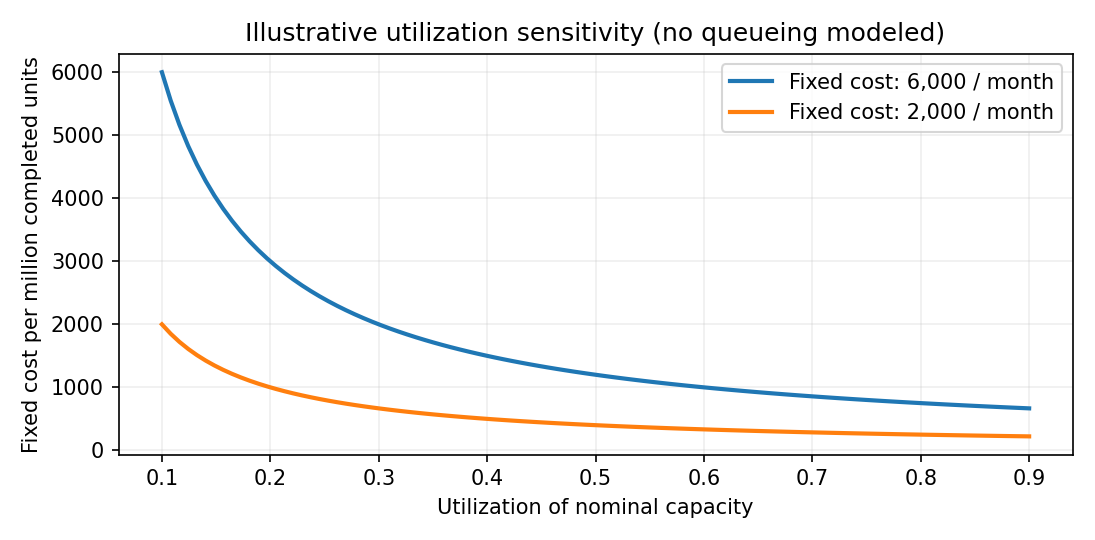}
\caption{Illustrative fixed-cost sensitivity to utilization, excluding queueing and reliability penalties.}
\label{fig:utilization}
\end{figure}

Nominal accelerator capacity is not the same as useful inference output. A server may have idle time because of demand seasonality, insufficient model memory, cold-start latency, queue fragmentation, maintenance, or privacy-policy restrictions. A fleet's aggregate theoretical throughput should not be treated as universally fungible capacity. If the monthly fixed cost is $F$, the nominal successful-workload capacity is $K$, and the realized useful utilization is $u$, fixed cost per completed unit is $F/(uK)$ (Figure~\ref{fig:utilization}). This expression captures the fixed-cost dilution but excludes queueing and failure; it is therefore a partial model.

The economically relevant utilization for distributed infrastructure is \emph{eligible utilization}: the fraction of capacity that can actually serve the model and data class in question. An idle GPU outside the approved jurisdiction has zero eligible capacity for a strictly local workload. Similarly, a device with insufficient memory for the required model cannot be credited with its advertised raw compute throughput.

\subsection{Counterfactual ownership accounting}
Two legitimate but distinct questions arise. \emph{Incremental deployment analysis} asks whether running an additional request on already-owned and otherwise idle hardware costs less than purchasing equivalent cloud inference. \emph{Full-resource analysis} asks whether society or an institution should own and operate that hardware in the first place. Incremental analysis can omit genuinely sunk purchase expenditure but must still include wear, energy, operational risk, and opportunity cost. Full-resource analysis includes capital amortization and allocated overhead. Reporting only the former while comparing against a fully loaded cloud tariff exaggerates the claimed system-wide advantage.

\section{Sovereignty and Trust as Formal Constraints}

\subsection{A data-flow-oriented threat model}
A privacy claim must specify the protected asset, attacker, trusted components, and observable surfaces. Assets include raw prompts, retrieved documents, model weights, embeddings, outputs, caches, telemetry, and usage receipts. Potential attackers include a compromised scheduler, malicious worker, privileged host administrator, external network observer, dishonest billing provider, and unauthorized downstream recipient. Different controls mitigate different threats: encrypted transport protects traffic in transit but not plaintext at the endpoint; region restrictions control geographic placement but do not prevent privileged access within the region; trusted execution can reduce host exposure only under a verified hardware and attestation model.

The minimum policy invariant is that sensitive request $i$ must never be dispatched to an execution domain outside its authorized set $A_i$. Formally, for every execution event $(i, j, t)$, require $j \in A_i$. This property can be tested through audit logs, fault injection, and attempts to trigger fallback routes. However, passing those tests demonstrates routing-policy enforcement under the tested conditions, not cryptographic confidentiality against a malicious approved worker.

\subsection{Four deployment trust regimes}
\emph{Institution-private inference} retains model execution within one organization's administrative boundary. \emph{Trusted regional consortium inference} allows execution at explicitly approved counterparties under shared governance. \emph{Open-market inference} permits untrusted operators for non-sensitive workloads, requiring strong isolation, integrity checks, and dispute procedures. \emph{Confidential-compute inference} adds hardware-based isolation and attestation where actually supported. These regimes are not a maturity ladder: a private institutional deployment may be the appropriate final state for highly sensitive information, even if an open marketplace offers a lower nominal price.

\subsection{Governance dependencies}
Sovereignty is weakened when a locally hosted model requires foreign authentication, model download, key escrow, telemetry collection, or an unavailable control-plane service to operate. The evaluation should therefore inventory software supply chains, identity providers, signing keys, administrative access, backup locations, and offline operating modes. A locality-only architecture is not necessarily an autonomous architecture. For public-sector and regulated institutions, a dependency graph may be as consequential as the location of the accelerator.

\section{Queueing, Heterogeneity, and Tail Performance}

\subsection{Why mean latency is insufficient}
Distributed fleets have variable network round-trip time, device performance, model warm-up, and competing workloads. A scheduler that minimizes mean completion time can still expose users to unacceptable p99 delays. The empirical objective should distinguish time to first token (TTFT), inter-token latency, total completion time, admission delay, and retry time. For interactive applications, TTFT and p95/p99 completion latency can be more decision-relevant than aggregate tokens per second.

\begin{figure}[htbp]
\centering
\includegraphics[width=0.72\textwidth]{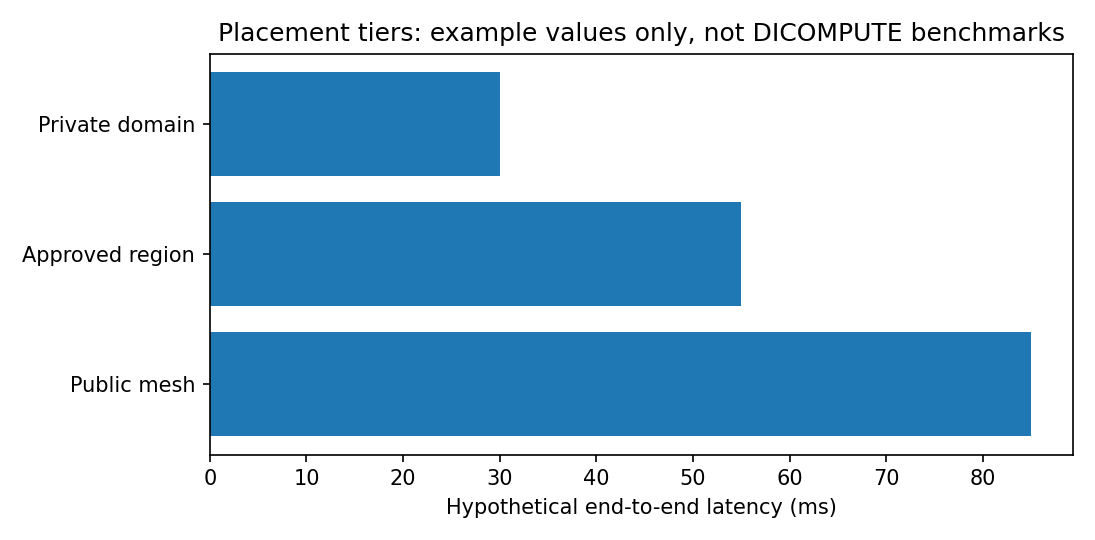}
\caption{Conceptual placement tiers. The numbers are hypothetical and only illustrate that trust policy and performance may vary by placement.}
\label{fig:tiers}
\end{figure}

A simple queueing approximation illustrates a structural issue: as service utilization approaches capacity, waiting time can grow sharply even if the device's raw token throughput remains unchanged. Heterogeneity complicates this because service times depend on model, sequence length, quantization, memory pressure, and batching. Consequently, a cost-optimal policy should enforce tail-latency constraints and may intentionally reserve headroom, increasing apparent cost per token while improving reliability (Figure~\ref{fig:tiers}).

\subsection{Replication and correlated failure}
Replicas improve availability only when failure modes are sufficiently independent. If replicas depend on the same identity service, coordinator, region, software update channel, or upstream model registry, nominal geographic distribution can mask a single point of failure. Evaluate isolated worker outages separately from shared coordinator failures, routing partition, authentication outage, and synchronized software defects. The measured quantity should be completed requests within SLO during disruption, not merely whether a process remained online.

\subsection{The bandwidth boundary}
Request-level distribution can be efficient when the prompt and output are modest relative to local computation and models are already resident on the worker. Layer-wise model parallelism across distant nodes, by contrast, requires transferring intermediate activations repeatedly and can be dominated by network latency and bandwidth. The paper therefore does not assume that a commodity wide-area mesh can replace high-bandwidth accelerator interconnects for tightly coupled large-model inference. This boundary is a central reason to treat distributed serving as a complement to, rather than a universal replacement for, centralized clusters.

\section{Research Hypotheses, Experimental Matrix, and Reporting Discipline}

The following hypotheses are proposed for preregistered testing, not asserted as proven findings.
\begin{itemize}
  \item \textbf{H1 (economic):} For selected request-level workloads and compatible underutilized hardware, policy-aware placement can reduce reliability-adjusted incremental inference cost relative to a matched centralized alternative.
  \item \textbf{H2 (sovereignty):} Under an enforced allowlist and tested fallback behavior, unauthorized cross-domain execution events can be reduced to zero in the specified test suite, with a measurable admission or cost penalty.
  \item \textbf{H3 (performance):} The relative advantage of a heterogeneous fleet decreases as context length, inter-node communication, or tail-latency strictness increases.
  \item \textbf{H4 (resilience):} Multi-domain placement improves completion under isolated worker failure but offers limited protection against shared control-plane failure.
  \item \textbf{H5 (accounting):} Off-critical-path settlement reduces settlement-induced response latency while creating a measurable reconciliation and dispute-management cost.
\end{itemize}

A useful factorial matrix crosses: two or more model sizes; short and long prompts; short and long generations; steady and bursty arrivals; private, regional, and open eligibility policies; low and high eligible utilization; and independent versus correlated faults. The matrix should be reduced through preregistered stratification if a full factorial experiment is impractical. The central requirement is matched conditions, not a large number of loosely comparable runs.

\begin{table}[htbp]
\centering
\small
\caption{Outcomes, measurements, and interpretation constraints.}
\label{tab:outcomes}
\begin{tabularx}{\textwidth}{L L L}
\toprule
Outcome & Measurement & Interpretation constraint \\
\midrule
Cost per SLO-compliant completion & Full cost ledger / successful eligible completions & Include retries and rejections \\
Input and output token economics & Separate rates for prefill and decode & Match model and sequence mix \\
p50 / p95 / p99 latency & End-to-end request traces & Include admission and cold starts \\
Policy violations & Authorized-domain audit events & Does not prove execution confidentiality \\
Energy per useful output & Metered energy / accepted outputs & Include idle and network allocation \\
Availability under faults & Successful completions during injected events & Report common-mode dependencies \\
Provider concentration & Workload share by failure domain & More nodes need not mean more independence \\
\bottomrule
\end{tabularx}
\end{table}

Report confidence intervals from repeated matched traces, and preserve negative results. Where performance data are unavailable, tables should be labeled \emph{not measured} rather than populated with simulated numbers. Simulation is appropriate for sensitivity analysis, but its parameters and source code must be disclosed. Public datasets must be checked for licensing and privacy restrictions; healthcare and banking examples should use approved synthetic data unless explicit consent and governance approvals exist.

\section{Positioning Against Existing Research}

The contribution proposed here is an integrated decision framework rather than a claim to have invented distributed inference. Orca and vLLM establish serving efficiency within accelerator deployments~\cite{orca,vllm}; DistServe and Splitwise explore prefill/decode disaggregation~\cite{distserve,splitwise}; Petals explores collaborative model execution~\cite{petals}; FlexGen addresses memory-constrained serving~\cite{flexgen}. DICOMPUTE's research question differs: how to place eligible inference requests across heterogeneous administrative domains while jointly accounting for trust, economic opportunity cost, latency, and auditability. The framework should be evaluated against these systems where their design goals overlap, without suggesting they are direct substitutes in every workload regime.

This distinction matters for academic novelty. A policy-aware placement formulation by itself may be viewed as an application of established scheduling and operations research. A stronger publishable contribution would be a validated model of the joint cost--trust--performance frontier, accompanied by reproducible traces and evidence that a practical policy improves a well-defined baseline. The paper therefore presents its mathematical framing as a research hypothesis and analytical synthesis until empirical novelty is demonstrated.

\section{Limitations and Scope of Claims}

The present manuscript is a theoretical research paper with illustrative sensitivity plots. It is not a report of a controlled DICOMPUTE hardware benchmark. It does not establish numerical cost savings, execution confidentiality on untrusted hosts, universally superior resilience, or a verified reduction in carbon emissions. All such claims require workload-matched measurements, an explicit threat model, and appropriate systems boundaries. This preprint has not been peer reviewed.

\section{Policy-Constrained Placement as a Feasibility Problem}

The design should begin with admissibility rather than optimization. Let $I$ denote requests and $J$ denote candidate execution sites. Each request $i$ has a permitted set of jurisdictions $R_i$, an approved set of administrative operators $O_i$, a required model and numerical precision $M_i$, and a latency budget $L_i$. Each site $j$ has jurisdiction $r_j$, operator $o_j$, model availability $m_j$, and available memory $v_j$. A binary decision $x_{ij}$ indicates whether request $i$ is assigned to site $j$. The feasible assignment set is
\begin{equation}
\mathcal{F}_i = \left\{ j \in J : r_j \in R_i,\; o_j \in O_i,\; m_j \succeq M_i,\; v_j \ge v_i,\; \hat{\ell}_{ij} \le L_i \right\}.
\end{equation}

The scheduler may minimize a weighted objective only over $\mathcal{F}_i$. If $\mathcal{F}_i$ is empty, rejecting the request is a valid and sometimes necessary result. This makes a privacy restriction operationally meaningful: no finite cost discount should override a hard confidentiality or residency rule. In contrast, cost, carbon intensity, and expected waiting time can be soft preferences when the institution permits trade-offs.

\begin{figure}[htbp]
\centering
\includegraphics[width=0.85\textwidth]{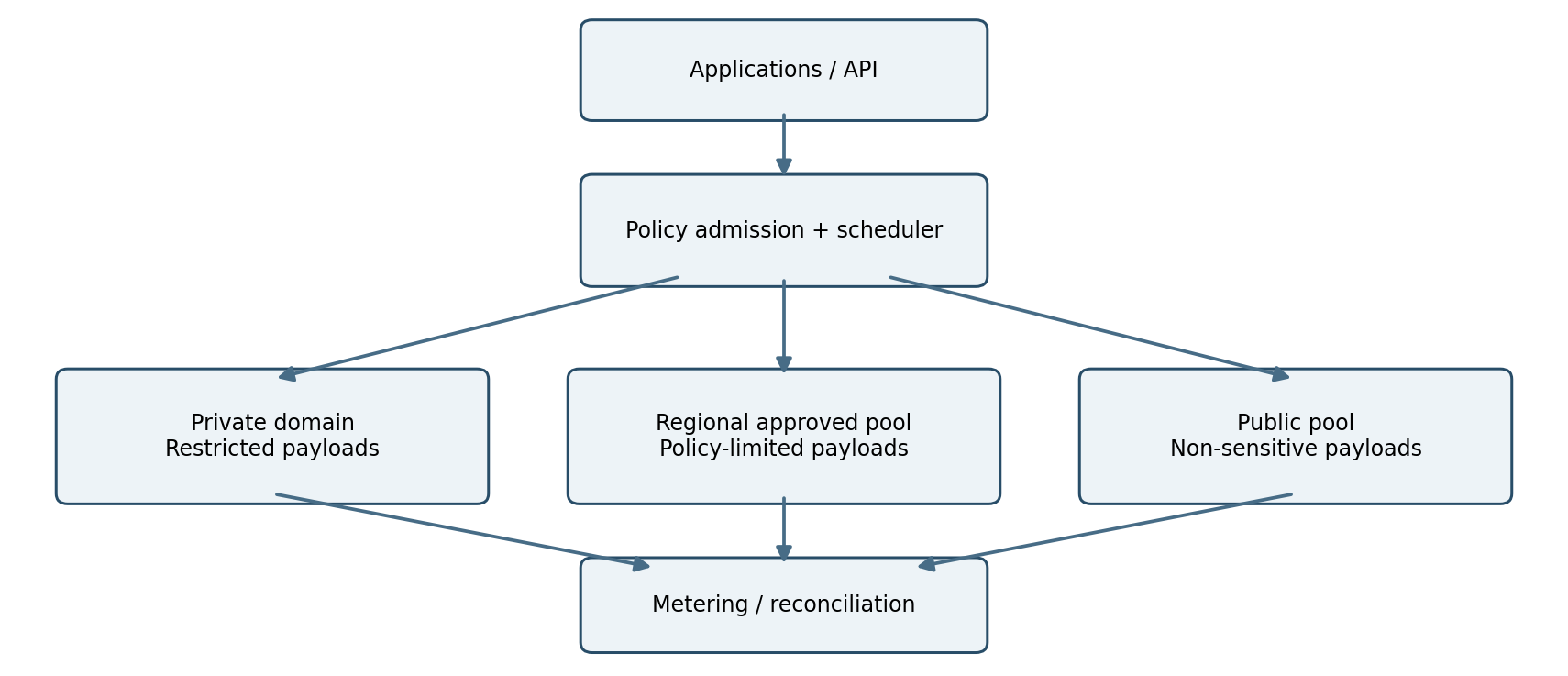}
\caption{Conceptual policy-gated inference architecture and administrative trust domains.}
\label{fig:policygated}
\end{figure}

A candidate optimization is
\begin{equation}
\min_{x} \sum_{i,j} x_{ij}\left(c_{ij} + \lambda_{\ell}\, \hat{\ell}_{ij} + \lambda_f\, p^{\mathrm{fail}}_{ij}\right),
\end{equation}
subject to $\sum_j x_{ij} \le 1$, $x_{ij} = 0$ for $j \notin \mathcal{F}_i$, and capacity constraints. The coefficients are institution-specific; their values must be reported rather than tuned after seeing results. This formulation is an analytical abstraction and is not claimed as a novel optimization algorithm. Figure~\ref{fig:policygated} summarizes the resulting policy-gated architecture.

\subsection{Locality does not imply confidentiality}
\begin{figure}[htbp]
\centering
\includegraphics[width=0.85\textwidth]{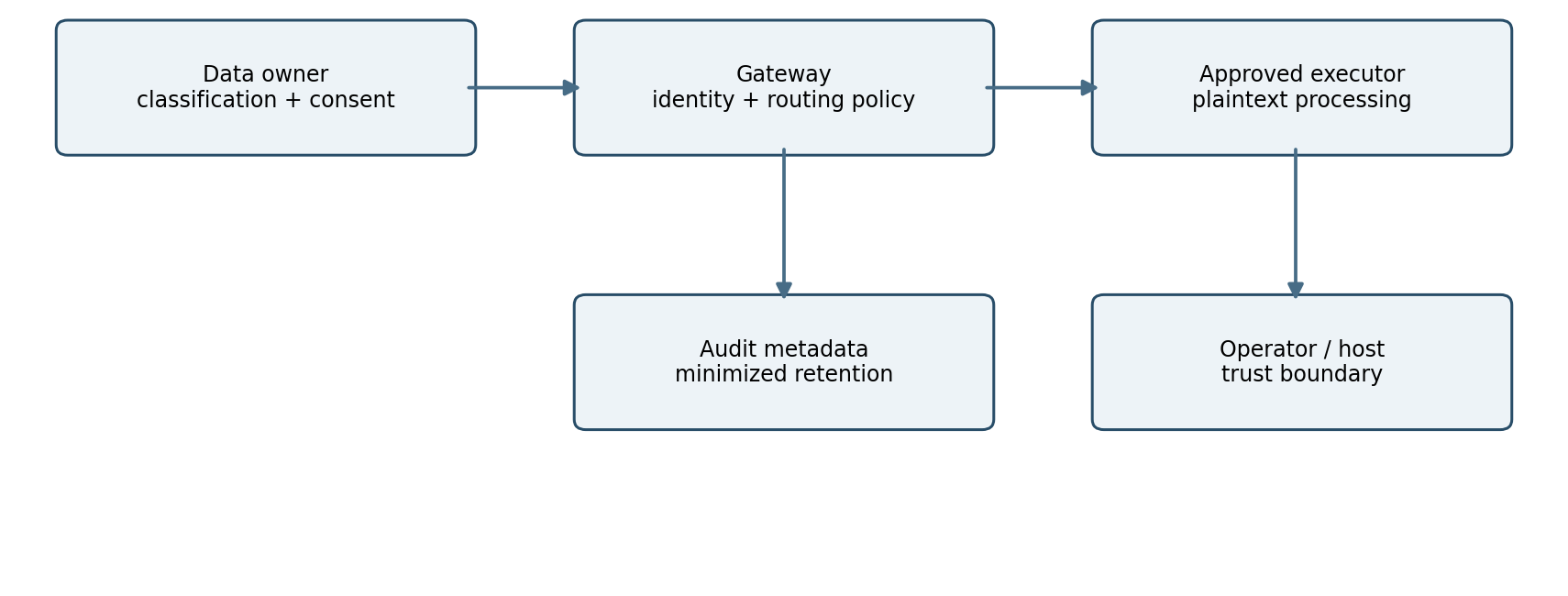}
\caption{Distinct privacy boundaries for request routing, execution, and audit metadata. Transport encryption is not execution secrecy.}
\label{fig:boundaries}
\end{figure}

An institution may keep data within its jurisdiction yet still expose plaintext to a cloud administrator, compromised hypervisor, diagnostic logging system, or external model provider. Data residency describes where processing occurs; administrative sovereignty describes who controls the relevant systems; cryptographic confidentiality describes what an adversary can learn. These dimensions are related but not interchangeable (Figure~\ref{fig:boundaries}). An honest evaluation should measure unauthorized dispatch separately from exposure to privileged personnel or compromised execution hosts.

\section{Economic Frontier and Decision Rules}

Let $Q$ be the number of requests meeting a fixed quality and service-level objective over a measurement period. Define reliability-adjusted unit cost as $U = (F + V + N + G + S + E)/Q$, where $F$ denotes amortized fixed infrastructure cost, $V$ variable compute, $N$ network transport, $G$ governance and operations, $S$ security and compliance, and $E$ expected cost of retries, failure recovery, and rejected requests. This expression must be evaluated with the same accounting horizon and allocation policy for each deployment. If capacity already exists, incremental cost may omit sunk acquisition expense, but the economic comparison should also report the opportunity cost of alternative uses and a full-cost case including depreciation.

\subsection{Three distinct economic questions}
\emph{Build-versus-buy:} Should an organization purchase hardware or purchase cloud inference? This depends on expected load duration, capital cost, support staffing, and residual value. \emph{Placement:} Given hardware already owned, where should an eligible request execute? This depends on the marginal cost of a successful completion and the current queue. \emph{Market participation:} Should an owner offer spare capacity to other institutions? This additionally depends on contractual risk, network exposure, tenant isolation, utilization variability, and payment collection. These questions cannot be answered by a single generic price-per-token estimate.

\subsection{When distribution loses}
A distributed fleet can lose its apparent capital advantage when low-powered devices require excessive replication, model loading dominates request time, requests have long contexts, egress fees are high, or policy constraints fragment the available capacity. Centralized clusters may also achieve materially higher utilization through batching and pooled demand. Consequently, the paper proposes reporting a frontier across load, model size, latency objective, and trust restrictions rather than a single savings percentage. A negative or null result for a workload is scientifically informative.

\subsection{Sensitivity reporting}
A publishable economic analysis should vary device lifetime, electricity price, cooling overhead, annual maintenance, utilization, bandwidth charges, failure probability, and eligible fleet fraction. Each assumption should be sourced or labeled hypothetical. Confidence intervals for measured throughput must propagate into unit-cost uncertainty. Because hardware generations and cloud tariffs change, the experimental artifact should store dated prices separately from performance observations so that economics can be recomputed without rerunning all inference tests.

\section{Evaluation Protocol and Reproducibility Standard}

We propose three baseline classes: a centralized cloud GPU deployment, a dedicated institutional cluster, and a heterogeneous policy-aware fleet. All should serve the same model family, quantization where possible, input/output distributions, and target output-quality criteria. If exact equivalence is infeasible, the paper must state the mismatch and avoid ranking the systems as directly comparable. Requests should include short interactive prompts, long-context generation, and asynchronous batch tasks. Report time to first token, tokens per second, p50/p95/p99 latency, energy per accepted output, cost per SLO-compliant completion, and policy-violation count.

\begin{figure}[htbp]
\centering
\includegraphics[width=0.85\textwidth]{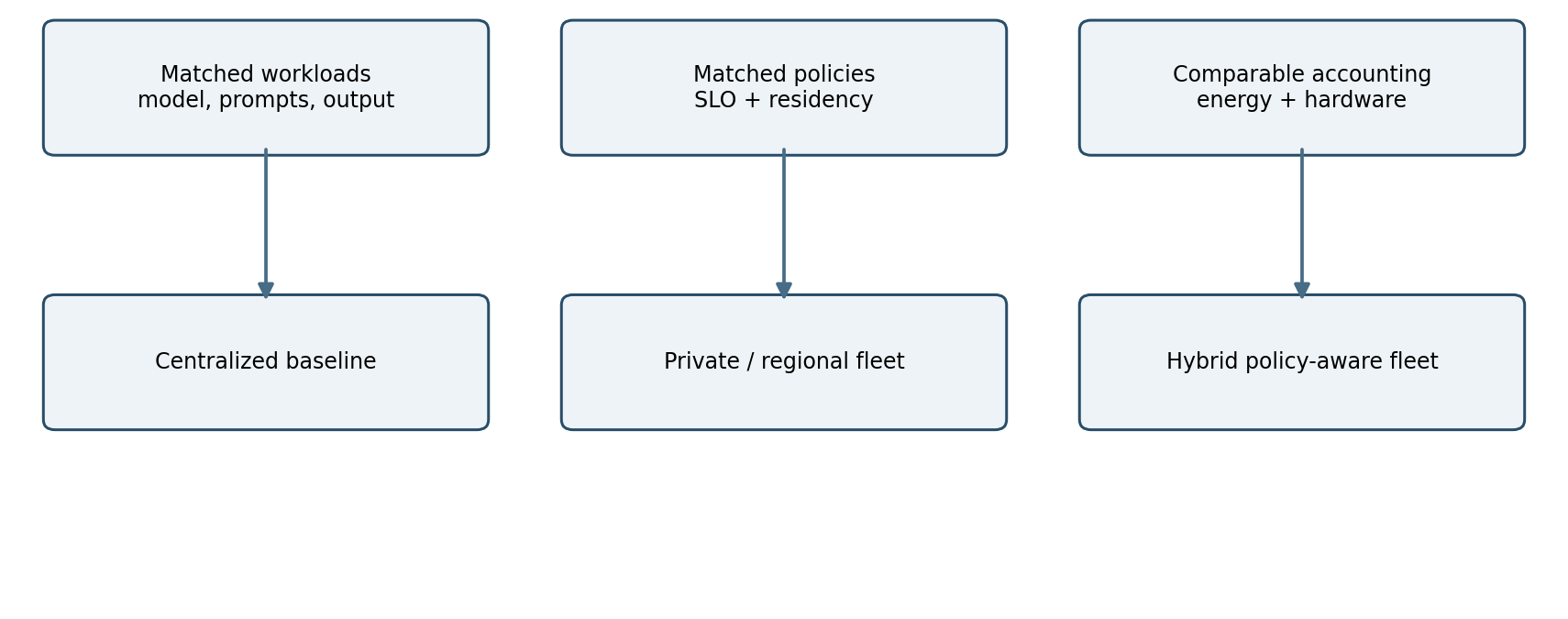}
\caption{Controlled comparison across baseline deployment classes (matched experimental design); no benchmark measurements are implied.}
\label{fig:baselines}
\end{figure}

For each workload, collect warm and cold-start observations, repeat trials on separate days, and use paired traffic traces where feasible (Figure~\ref{fig:baselines}). Pre-register the primary endpoint before selecting the best configuration. Measure the effect of worker churn, an unreachable coordinator, and a shared authentication outage. Distinguish node-level fault tolerance from system-level availability. For privacy tests, inject requests tagged with restrictive residency rules and test whether fallback behavior ever dispatches them to prohibited domains; a zero-violation test result is limited to the tested threat and failure model.

A reproducibility bundle should include a hardware inventory, power measurement method, model identifiers and licenses, serving runtime versions, anonymized request traces, scheduler configuration, tariff assumptions, and scripts that regenerate figures. No patient or banking customer data should be published in the benchmark corpus. Where sensitive workloads motivate the study, use approved synthetic analogues and document how their token-length and burstiness distributions were selected.

\section{Scholarly Positioning and Scope of Claims}

The proposed contribution is a joint cost--trust--performance decision framework for inference across administrative domains. It should not be framed as inventing edge computing, privacy-preserving machine learning, continuous batching, or decentralized inference. Existing systems research already establishes sophisticated scheduling, memory management, and disaggregated serving approaches. DICOMPUTE's potential distinction is the treatment of economic opportunity cost and institutional policy as first-class placement constraints, with audit and settlement logically separated from the request's latency-critical path.

This distinction is a research proposition, not proof of priority over the literature. The framework supports a theoretical contribution without production benchmarks because the abstract and conclusions label analytical and illustrative findings as such; a strong systems-evaluation claim requires reproducible measured results.

\section{Conclusion}

Distributed inference is best understood as an \emph{additional operating point} in the AI infrastructure design space, not a categorical replacement for centralized data centers. DICOMPUTE motivates a research program in which placement is governed jointly by cost, performance, trust, locality, and verifiable economic coordination. The central theoretical result is conditional: underutilized compatible resources, policy-constrained workloads, and sufficiently low orchestration overhead can make distributed serving attractive, while high-utilization clusters and tightly coupled models preserve strong centralized advantages. Publication-quality empirical work should test these boundaries, report uncertainty, and state privacy guarantees precisely.

\section*{Competing Interests}
The authors are members of the XDC Network research and engineering team. XDC is discussed in Section~\ref{sec:metering} as one possible optional settlement or audit-anchoring layer; no claim in this paper depends on the use of any particular blockchain.


\appendix
\section{Reporting Template for Future Empirical Results}

For each run, publish the model identifier and checksum, hardware CPU/GPU/RAM, serving engine and version, quantization, batch size, concurrency, prompt distribution, input and output token counts, energy-meter methodology, cost allocation assumptions, trust policy, and raw trial outputs. Include reproducible scripts and anonymized traces when permitted. Separate measured quantities from modeled estimates and scenario assumptions in all tables and figures.

\end{document}